\documentclass{elsart5p}

\usepackage[bookmarks=true, bookmarksopen=true]{hyperref}

 \usepackage{graphicx}

\usepackage{amssymb}

\begin{document}

\begin{frontmatter}



\title{Multiband and impurity effects in infrared and optical spectra of MgB$_{2}$}

\author{A.B. Kuzmenko}

\address{DPMC, University of Geneva, 1211 Geneva 4,
Switzerland}

\begin{abstract}

A short review of the optical and far-infrared measurements on
MgB$_{2}$ is given. Multiband and multigap effects are analyzed
by comparing optical properties with other experiments and {\em
ab initio} calculations. The covered topics are: the plasma
frequency, electron-phonon interaction, impurity scattering,
the effects of C and Al substitution, interband transitions and
the far-infrared signatures of the superconducting gaps.

\end{abstract}

\end{frontmatter}

\section{Introduction}
\label{Intro}

The existence of two distinct superconducting gaps in a
combination with a comfortably high $T_{c}$ of 40 K in MgB$_{2}$ offers
a unique opportunity to study the electrodynamics of a multigap
superconductor. The effects of interband coupling and
scattering on the superconducting order parameter and $T_{c}$
were envisaged long time ago \cite{SuhlPRL59}, but only after
the discovery of superconductivity in magnesium diboride
\cite{NagamatsuNature01} the community was given a real chance
to study the phenomenon experimentally. MgB$_{2}$ seems to be a
very unusual case of 'conventional' superconductivity, where
the novel multiband and multigap physics can be well captured
by {\em ab initio} calculations of the electron and phonon
dispersions, electron-phonon interaction and properly modified
strong coupling Eliashberg formalism \cite{MazinPC03}. As a
result, a rather direct quantitative comparison between theory
and various experimental probes is plausible, in contrast to
the case of high-$T_{c}$ cuprates with an important role of
electron correlations and still debated mechanism of
superconductivity.

Electronic band structure is the starting point to understand
the unusual superconducting properties of MgB$_{2}$. The
first-principles calculations that exist since the late 1970's
\cite{TupitsynSPSS76} were refined after 2001 by several groups
\cite{KortusPRL01,KongPRB01,AnPRL01,RavindranPRB01,ChoiPRB02}.
Angle-resolved photoemission (ARPES) \cite{UchiyamaPRL02} and
de Haas-van Alphen (dHvA) measurement \cite{YellandPRL02}
showed a good agreement with the calculated band dispersion.
The two-gap behavior, first predicted theoretically
\cite{LiuPRL01}, was confirmed by tunneling spectroscopy
\cite{IavaronePRL02,EskildsenPRL02}, specific heat
\cite{WangPC01}, ARPES \cite{SoumaNature03}, Raman
\cite{QuiltyPRL02} and other techniques.

Until recently, the agreement of optical data with theoretical
predictions has been far less successful. Perhaps, the most
notable and serious issue was an extremely small plasma
frequency reported from several early optical studies
\cite{TuPRL01,KuzmenkoSSC02,MunJS02,FudamotoPRB03,ChvostovaTSF04}.
More recent studies
\cite{GuritanuPRB06,DiCastroPRB06,KakeshitaPRL06} showed that
the plasma frequency is actually much closer to the expected
value, although optics results call for certain amendments to
the theory. The main problem in obtaining reasonably
reproducible optical data is the very small size of the
existing high-quality single crystals
\cite{KarpinskiSST02,LeePC03}.

The electron-phonon interaction, which is responsible for the
superconductivity in MgB$_{2}$ as suggested by the isotope
effect \cite{BudkoPRL01,HinksNature01}, inevitably manifests
itself via the energy dependent renormalization of the
scattering rate and the effective mass. The latter parameters
can be derived from the optical spectra already in the normal
state, provided that the correct plasma frequency is available.
Single crystal measurements \cite{GuritanuPRB06} have shown a
reasonably good agreement with the {\em ab initio} calculations
\cite{KongPRB01,LiuPRL01,ChoiPRB02} and confirmed a stronger
electron-phonon coupling in the $\sigma$ bands.

While the biggest controversies regarding the general
normal-state optical properties of MgB$_{2}$ seem to be
settled, the main issue related to far-infrared spectra in the
superconducting state, namely the absence of a clear evidence
of the two-gap structure, still remains. Unlike the situation
with early optical measurements on magnesium diboride, a number
of different far-infrared experiments, especially those on thin
films, are quite consistent with each other. Therefore we have
to ask ourselves whether the existing theory describes
adequately the infrared optical response of a two-gap
superconductor.

The organization of this paper is as follows: In Section
\ref{Electrodynamics} some basic considerations regarding the
multiband conductivity are given. Section \ref{Plasma} deals
with the experimental determination of the plasma frequency. In
Section \ref{Scattering} the electron-phonon and impurity
scattering as observed by optics are discussed. Interband
transitions are considered in Section \ref{Interband}. Finally,
the far-infrared signatures of the superconducting gap(s) are
reviewed in Section \ref{FIR}.

\section{Conductivity of a multiband system}
\label{Electrodynamics}

The two main contributions to the complex optical conductivity
$\sigma(\omega)=\sigma_{1}(\omega)+i\sigma_{2}(\omega)$ of a
metal are the intraband (Drude) component centered at zero
frequency and the one due to momentum conserving interband
transitions
\begin{eqnarray}\label{SigDrudeIB}
\sigma(\omega) = \sigma_{D}(\omega)+\sigma_{IB}(\omega).
\end{eqnarray}

\noindent The detailed theory of multiband transport based on
the Boltzmann equation was developed in Ref.\cite{PinskiPRB81}.
A widely used approximation for the Drude conductivity
$\sigma_{D}(\omega)$ of a system with several conduction bands
is the parallel-resistors formula, where each resistor (or
better say conductor) corresponds to a separate band. Although
this seems like assuming that quasiparticles in different bands
do not 'see' each other, the impurity or phonon induced
interband scattering probabilities to a certain approximation
can be just added to the intraband ones, resulting in a
parallel band conduction with effectively renormalized
scattering rates.

The metallic properties of MgB$_{2}$ are determined by two
distinct types of electronic bands coming  almost entirely from
the boron states: the strongly covalent almost two-dimensional
$\sigma$ bands formed by hybridized $sp_{x}p_{y}$ orbitals and
the more isotropic $\pi$ bands made of $p_{z}$ orbitals. The
holes in the $\sigma$ bands are strongly coupled to the
in-plane $E_{2g}$ phonon mode at $\sim$ 75 meV, giving rise to
a high electron-phonon coupling constant
\cite{KongPRB01,LiuPRL01,ChoiPRB02}. In this paper we shall
make no distinction between bands of the same type,
representing the conductivity by the sum of the two terms

\begin{eqnarray}\label{SigSum}
\sigma_{D}(\omega)
=\sum_{\beta=\sigma,\pi}\sigma_{D,\beta}(\omega).
\end{eqnarray}

\noindent The optical sum rule reads as follows
\begin{eqnarray}\label{PlSum}
\int_{0}^{\infty}\sigma_{1D}(\omega)d\omega
=\frac{1}{8}\sum_{\beta=\sigma,\pi}\omega_{p,\beta}^2=\frac{1}{8}\omega_{p}^2,
\end{eqnarray}
\noindent where $\omega_{p,\sigma}$ and $\omega_{p,\pi}$ are
the unscreened (bare) plasma frequencies of the $\sigma$ and
$\pi$ bands and $\omega_{p}$ is the total plasma frequency. Of
course, the conductivity and the plasma frequency are both
tensors with two different components: one parallel to the ab
plane and another along the c axis. The first-principle
calculations predict the $\sigma$ band plasma frequency to be
very small along the c axis.

It follows from Eq.(\ref{SigSum}) that there is no rigorous
experimental way to separate $\sigma_{D,\sigma}(\omega)$ and
$\sigma_{D,\pi}(\omega)$. However, the multiband structure may
become apparent in optical spectra due to the strong disparity
between the two bands. Such a contrast can stem from (i)
different anisotropies of the plasma frequencies
$\omega_{p,\sigma}$ and $\omega_{p,\pi}$, (ii) manifestly
different electron scattering on phonons and impurities, and
(iii) a large difference of superconducting gap values. In the
following sections these possibilities will be analyzed in
details.

\section{Plasma frequency}
\label{Plasma}

The plasma frequency that can be measured using the optical sum
rule (\ref{PlSum}) is determined by the electronic band
dispersion $\epsilon_{k\beta}$ \cite{LiuPRL01}:
\begin{eqnarray}\label{PlFr}
\frac{\omega_{p,\beta\alpha}^{2}}{8}=\frac{\pi e^2}{\hbar^2
V}\sum_{k}\left(\frac{\partial \epsilon_{k\beta}}{\partial
k_{\alpha}}\right)^{2} \left.\left(-\frac{\partial f}{\partial
\epsilon}\right)\right|_{\epsilon=\epsilon_{k\beta}}\mbox{, \ }
(\alpha=a,c)
\end{eqnarray}

\noindent $V$ is the sample volume, $f$ is the Fermi-Dirac
distribution function and $\beta$ is the band index. At not too
high temperatures, only states close to the Fermi energy
$E_{F}$ contribute to Eq.(\ref{PlFr}), and $\omega_{p}^{2}\sim
N(E_{F})\langle v_{F}^{2}\rangle$, where $N(\epsilon)$ is the
density of states and $v_{F}$ is the Fermi velocity.

The plasma frequency is an important parameter for testing the
consistency of band structure calculations with optics.
Therefore, the reported value of only 1.5 - 3.0 eV for
$\omega_{p}$ by several groups
\cite{TuPRL01,KuzmenkoSSC02,MunJS02,FudamotoPRB03,ChvostovaTSF04}
soon after the discovery of superconductivity in MgB2, as
compared to the theoretical prediction of 7 eV in both
directions, was rather puzzling. The situation changed
recently, when three independent measurements on high-quality
single crystals
\cite{GuritanuPRB06,DiCastroPRB06,KakeshitaPRL06} showed the
values of $\omega_{p}$ much closer to the calculated ones.

Given the large spread of optical data, one should critically
consider the experimental issues. Possible complications are
(i) sample purity, (ii) surface contamination, (iii) analysis
ambiguity due to a limited spectral range and (iv) an overlap
between the Drude peak and interband peaks.

Impurities in small amounts are not expected to affect
significantly the plasma frequency. However, they increase the
elastic scattering and shift the Drude spectral weight to
higher frequencies. Interestingly, in MgB$_{2}$ impurities can
affect differently the charge scattering in $\sigma$ and $\pi$
bands \cite{MazinPRL02}.

As observed by several groups, the optical quality of the
surface of MgB$_{2}$ poly- and single crystals degrades in air
within minutes after polishing
\cite{KuzmenkoSSC02,FudamotoPRB03,GuritanuPRB06,KakeshitaPRL06}.
To the author's knowledge, there is no information about the
surface degradation of MgB$_{2}$ films. The surface
contamination strongly reduces the absolute reflectivity,
pushing down the apparent value of the plasma frequency.

A source of inaccuracy not to be neglected is a set of
assumptions taken to experimentally derive and integrate
$\sigma_{1}(\omega)$. In particular, the error bars due to
high- and low-frequency extrapolations in the Kramers-Kronig
(KK) analysis of reflectivity that are hard to estimate may
affect significantly the value of $\omega_{p}$, especially if
spectra are available only up to few electronvolts. This
uncertainty is reduced if the ellipsometric technique is used
allowing the direct measurement of $\sigma_{1}(\omega)$ and
$\sigma_{2}(\omega)$.

Tu {\em et al.} \cite{TuPRL01} measured the reflectivity of
high-quality oriented films ($T_{c}$=39.6 K) with the c axis
perpendicular to the surface in the range from 3.5 meV to 2.7
eV and extracted $\sigma(\omega)$ by the KK transform. A value
of $\omega_{p}$ $\approx$ 1.8 eV was deduced from the sum-rule
analysis. Notably, the film reflectivity is very high at low
frequencies (compatible with the predicted high metallicity of
MgB$_{2}$) but becomes quite low above 0.5-0.7 eV as compared
to the recent data on single crystals
\cite{GuritanuPRB06,DiCastroPRB06,KakeshitaPRL06}. The same
trend of high-frequency reflectivity suppression is present in
the film study by Mun {\em et al.} \cite{MunJS02} and, to a
larger extent, by Chvostov\'{a} {\em et al.}
\cite{ChvostovaTSF04}. It is not clear at the moment whether
this is due to surface contamination, film strains or other
reasons.

\begin{figure}[htb]
\centerline{\includegraphics[width=8cm]{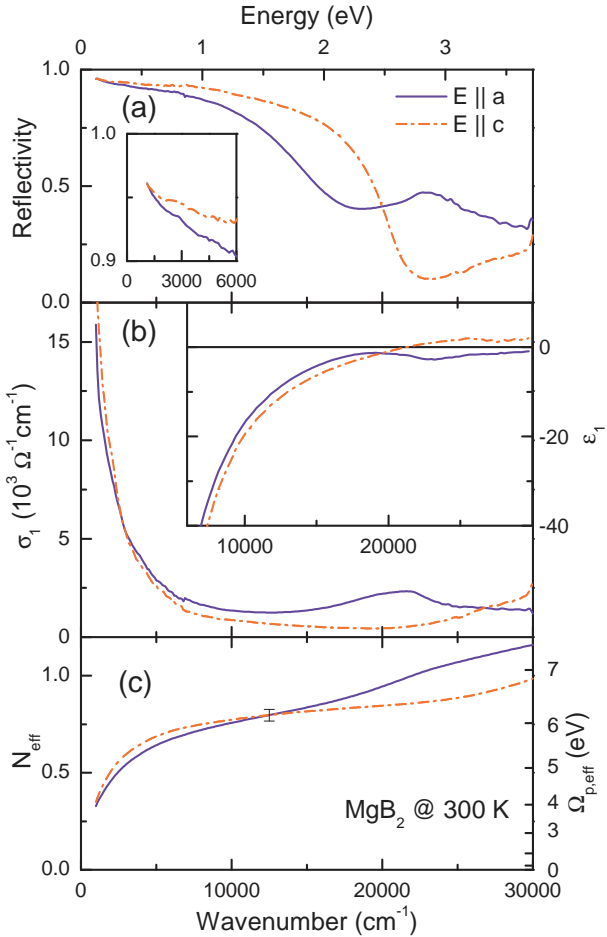}}
\caption{Optical anisotropic spectra of MgB$_{2}$ at 300 K
derived from ellipsometry and reflectivity measurements on an
ac-oriented single crystal: the normal-incidence reflectivity
$R(\omega)$(a), optical conductivity $\sigma_{1}(\omega)$, the
dielectric function $\epsilon_{1}(\omega)$ (b) the effective
number of carriers $N_{\mbox{\scriptsize eff }}(\omega)$ and
the effective plasma frequency $\Omega_{\mbox{\scriptsize p
eff}}(\omega)$ as a function of the integration cutoff energy
(c). Adapted from Guritanu {\em et al.} \cite{GuritanuPRB06}.
}\label{FigDataAC}
\end{figure}

In another early study, Kuzmenko {\em et al.}
\cite{KuzmenkoSSC02} obtained a direction-averaged conductivity
of a dense polycrystalline sample ($T_{c}$=39 K) from 6 meV to
4.6 eV by a combination of reflectivity and ellipsometry
measurements. The conductivity shows a narrow Drude peak with a
plasma frequency of only 1.4 eV and additionally a broad
infrared continuum with a spectral weight corresponding to the
plasma frequency of about 5 eV. Since this sample has a
sizeable contamination of MgO, a possible explanation of the
apparently small Drude plasma frequency is that the narrow peak
is due to the $\sigma$ bands, which are not strongly affected
by impurities, but renormalized by the electron-phonon
interaction, while the broad continuum is largely formed by the
$\pi$ bands \cite{MazinPRL02}. This is probably the main reason
for the low reflectivity of that sample, although some
influence of the surface contamination cannot be excluded.

The first attempt to distinguish $\omega_{p,a}$ and
$\omega_{p,c}$ was undertaken by Fudamoto and Lee
\cite{FudamotoPRB03}. From a comparison of the reflectivity
spectra measured on a mosaic of ab-plane oriented crystals and
on a polycrystalline sample they correctly deduced that the
plasma edge parallel to the ab plane is at about 2 eV while the
one along the c axis is at 2.75 eV. However, the conclusion of
Ref.\cite{FudamotoPRB03} about a significant difference between
$\omega_{p,a}$ and $\omega_{p,c}$ was not confirmed by the
latest single-crystal studies
\cite{GuritanuPRB06,KakeshitaPRL06}. The reason is that the
position of the plasma edge corresponds to the screened plasma
frequency
$\omega_{p}^{*}\approx\omega_{p}/\sqrt{\epsilon_{\infty}}$
which depends not only on $\omega_{p}$ but also on the
dielectric constant $\epsilon_{\infty}$ from interband
transitions. The latter quantity appears to be strongly
anisotropic in MgB$_{2}$ (Section \ref{Interband}).

A reliable determination of the optical anisotropy requires
single crystals with the c axis parallel to the surface. The
major difficulties are the small dimensions of the existing
single crystals, especially along the c axis ($<$ 200 $\mu$m)
and the mentioned surface degradation. Guritanu {\em et al.}
\cite{GuritanuPRB06} succeeded in doing spectroscopic
ellipsometry in the range 0.75 - 3.7 eV and reflectivity from
0.1 to 0.85 eV on the ac- and ab-surfaces of small single
crystals ($T_{c}$ = 38 K) at room temperature
(Fig.\ref{FigDataAC}), by adding focusing lenses to a standard
ellipsometry setup and keeping samples in a nitrogen flow. Both
in-plane and c-axis spectra exhibit a metallic behavior,
characterized by a reflectivity plasma edge, a Drude peak in
$\sigma_{1}(\omega)$ and a negative $\epsilon_{1}(\omega)$.
However, one can see a strong anisotropy of the plasma edge,
consistent with the assignment made in
Ref.\cite{FudamotoPRB03}. The plasma frequency derived by the
sum rule is about 6.3 eV for both directions, if the
integration is continued up to 1.7 eV. This cutoff value is
high enough to comprise most of the intraband spectral weight,
but it is below the interband peak at 2.6 eV along the a axis
(discussed in Section \ref{Interband}). Since ellipsometry
provides directly both real and imaginary parts of the
dielectric function, the value of the plasma frequency is
practically independent on extrapolations. The value of 6.3 eV
is much closer to 7 eV predicted by the LDA calculations. A
very small anisotropy of the plasma frequency is also in a good
agreement with the theoretical predictions.

\begin{figure}
\centerline{
\includegraphics[width=7.5cm]{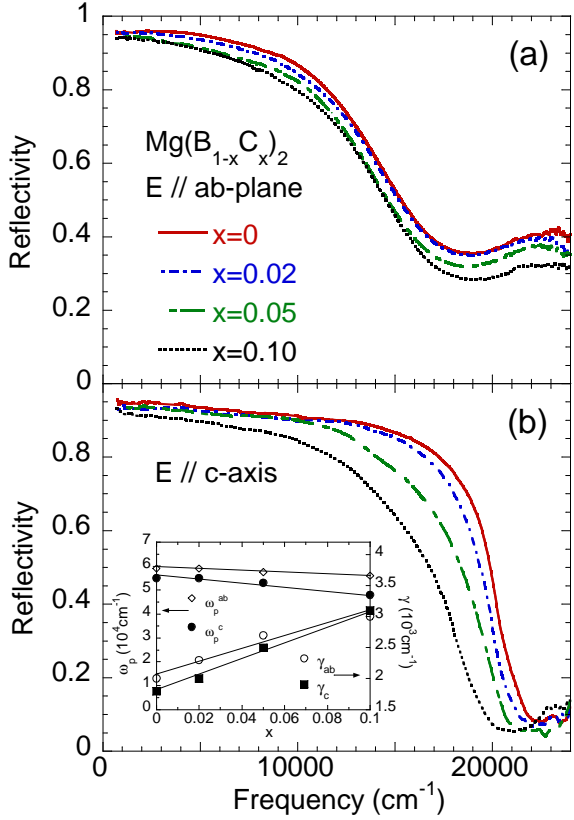}}
  \caption{The reflectivity spectra of Mg(B$_{1-x}$C$_{x}$)$_{2}$
for $E$ $\parallel$ $ab$ plane (a) and $E$ $\parallel$ $c$ axis
(b). Inset: the $x$ dependence of the plasma frequencies
$\omega_{p,a}$ and $\omega_{p,c}$, and the scattering rates
$\gamma_{a}$ and $\gamma_{c}$ obtained by a Drude-Lorentz fit.
Adapted from Kakeshita {\em et al.} \cite{KakeshitaPRL06}.
}\label{KakeshitaPRL06_Fig1}
\end{figure}

\begin{figure}
\centerline{
\includegraphics[width=8.5cm]{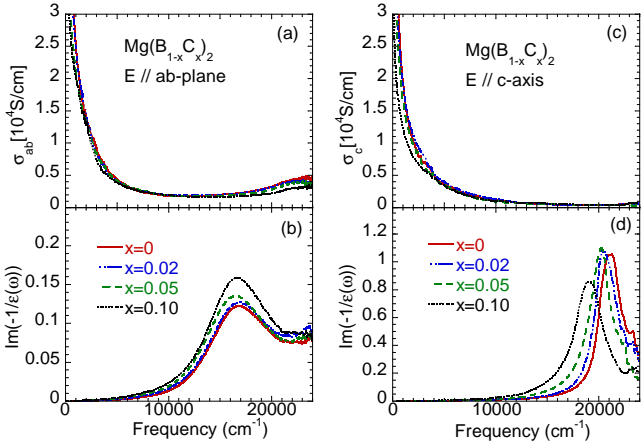}}
  \caption{The optical conductivity spectra [(a),
  (c)] and the loss function [(b),(d)] of
  Mg(B$_{1-x}$C$_{x}$)$_{2}$ for $E \parallel$ab plane and c
  axis obtained by the KK analysis of reflectivity curves from Fig.\ref{KakeshitaPRL06_Fig1}.
  Adapted from Kakeshita {\em et al.} \cite{KakeshitaPRL06}.
}\label{KakeshitaPRL06_Fig2}
\end{figure}

Independently, Kakeshita {\em et al.} \cite{KakeshitaPRL06}
measured a- and c-axis reflectivities of MgB$_{2}$ from 75 meV
to 3 eV on ac- and ab-oriented single crystals at room
temperature using a microscope spectroscopy technique
(Fig.\ref{KakeshitaPRL06_Fig1}). The sample was polished prior
to each measurement in order to avoid the effect of the surface
degradation. The data from Refs.\cite{KakeshitaPRL06} and
\cite{GuritanuPRB06} match rather well with each other,
although the plasma edges are sharper in
Ref.\cite{KakeshitaPRL06}, indicating a narrower Drude peak
(Fig.\ref{KakeshitaPRL06_Fig2}). The plasma frequencies
estimated in Ref.\cite{KakeshitaPRL06} from the optical sum
rule with a cutoff at 2.2 eV are $\omega_{p,a}$ $\sim$ 8.4 eV
and $\omega_{p,c}$ $\sim$ 7.0 eV. However, for such a high
cutoff the interband transition at 2.6 eV contributes
significantly to the sum rule along the a axis. A Drude-Lorentz
fit that treats the interband peak as a separate contribution
gives a smaller value $\omega_{p,a}$ $\sim$ 7.4 eV. The values
of the plasma frequencies obtained in Refs.\cite{GuritanuPRB06}
and \cite{KakeshitaPRL06} seem to agree each other within the
error bars of the KK analysis. In both cases, $\omega_{p,a}$
might be slightly overestimated since the theory predicts the
existence of a weak $\sigma\rightarrow\sigma$ interband
transition at 0.3-0.5 eV. Within the error margins, there is no
prohibitive disagreement between optics and LDA calculations,
as has been shown by independent measurements on single
crystals grown in different groups.

\begin{figure}[htb]

  \centerline{ \includegraphics[width=9cm]{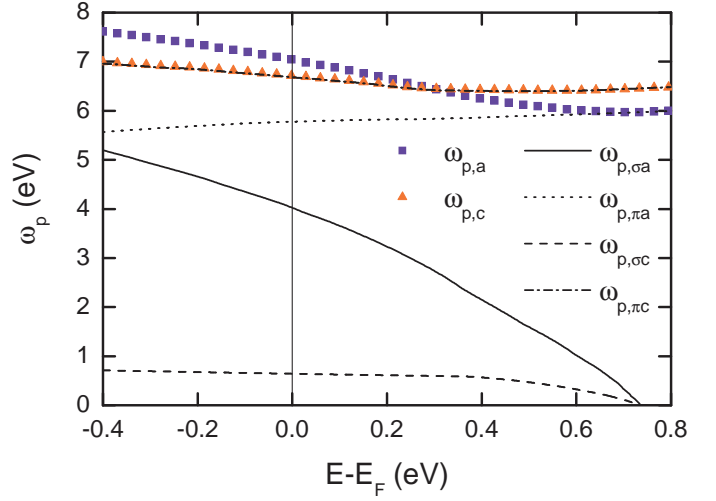}}
  \caption{Anisotropic plasma frequencies of $\sigma$ and $\pi$ bands as a function of the
  Fermi level in a rigid-band approximation.
  Also shown are the total plasma frequencies $\omega_{p,a}$
  and $\omega_{p,c}$. The data are taken from Ref. \cite{MazinPC03}.}\label{Wp}
\end{figure}

The influence of Al, C and other substitutions on the
properties of MgB$_{2}$ has been recently the focus of intense
research. All substitutions are found to suppress $T_{c}$
(\cite{KortusPRL05} and references therein). It was argued
\cite{KortusPRL05} that this suppression can be well described
by a combination of the filling of $\sigma$ bands and interband
scattering. Optics contributes to a better understanding of the
substitution effects. While Al replaces Mg and carbon
substitutes boron atoms, both of them dope electrons to the
system, which is expected to affect the plasma frequency. In
the rigid band picture, where the doping merely shifts the
Fermi level, the change of the band plasma frequencies can be
derived from the band structure calculations for the pure
compound. Fig.\ref{Wp} shows the plasma frequencies of $\sigma$
and $\pi$ bands as well as the total ones for the two
polarizations \cite{MazinPC03}. With the increase of the Fermi
level, both $\omega_{p,a}$ and $\omega_{p,c}$ are expected to
diminish, but the decrease of $\omega_{p,a}$ is faster due to a
rapid decrease of $\omega_{p,\sigma a}$. The latter effect is
caused by a fast depletion of the density of states (DOS) of
$\sigma$ bands.

Kakeshita {\em et al.} \cite{KakeshitaPRL06} measured
reflectivity on a series of carbon doped samples
Mg(B$_{1-x}$C$_{x}$)$_{2}$, $0 \leq x \leq 0.1$
(Fig.\ref{KakeshitaPRL06_Fig1}). Remarkably, the c-axis plasma
edge as well as the corresponding peak of the loss function
(Fig.\ref{KakeshitaPRL06_Fig2}) shift to low frequencies with
doping while the one for the a axis is almost doping
independent. Correspondingly, $\omega_{p,c}$ notably goes down,
while $\omega_{p,a}$ shows only a minor decrease with C
substitution (shown as inset of Fig.\ref{KakeshitaPRL06_Fig1}).
The decrease of both $\omega_{p,a}$ and $\omega_{p,c}$ with
doping qualitatively agrees with their calculated Fermi-level
dependence (Fig.\ref{Wp}). However, the rigid band
approximation does not explain that $\omega_{p,c}$ goes down
much faster than $\omega_{p,a}$. It is necessary therefore to
take into account the doping induced band modification. The
calculations of Refs. \cite{KortusPRL05} and
\cite{ProfetaPRB03} indeed show a much slower decrease of the
$\sigma$-band DOS compared to the rigid band model. This
results in a slower rate of the suppression of $T_{c}$ with
doping, in agreement with observations.

\begin{figure}[htb]
  \centerline{\includegraphics[width=9cm]{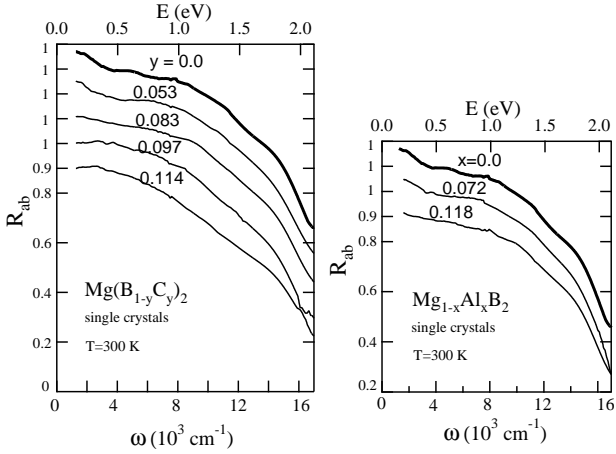}}
  \caption{In-plane reflectivity spectra of C- and Al-doped
single crystals of MgB$_{2}$. Adapted from Di Castro {\em et
al.} \cite{DiCastroPRB06}. }\label{DiCastroPRB06}
\end{figure}

The ab-plane optical reflectivity of a series of C- and
Al-doped single crystals of MgB$_{2}$ was measured by Di Castro
{\em et al.} \cite{DiCastroPRB06} (Fig.\ref{DiCastroPRB06}).
The data indicate that each of these substitutions causes only
a minor shift of the a-axis plasma edge. Correspondingly, a
quantitative analysis showed a slight decrease of the total
in-plane plasma frequency. As far as carbon doping is
concerned, this agrees well with the results of Ref.
\cite{KakeshitaPRL06}. In Ref. \cite{DiCastroPRB06} a value of
$\omega_{p,a}$ of about 5 eV was reported, which is somewhat
smaller than the values obtained in Refs.\cite{GuritanuPRB06}
and \cite{KakeshitaPRL06}. The low-frequency reflectivity is
close to the one from \cite{GuritanuPRB06}, but drops faster
above 2 eV, which partially explains, together with
uncertainties due to a limited spectral range, the difference
of the plasma frequencies. Another reason is that the model
dielectric function used by the authors \cite{DiCastroPRB06} to
fit reflectivity contained an extra mid-infrared (MIR) peak
ascribed to the $\sigma\rightarrow\sigma$ interband transition.

\section{Electron-phonon and impurity scattering}
\label{Scattering}

The information about electron-phonon interaction and impurity
scattering can be derived from the shape of the Drude peak. In
particular, the strength of the electron-phonon coupling, which
is widely believed to be responsible for the superconducting
pairing, can be deduced from the infrared spectra. The
knowledge of impurity scattering rates can be used to test the
hypothesis about a strong disparity between $\sigma$ and $\pi$
bands and a very small interband scattering used to explain the
survival of two distinct superconducting gaps
\cite{MazinPRL02}.

The broadening is conveniently described via the memory
function $M(\omega,T)=1/\tau(\omega,T)-i\omega
\lambda(\omega,T)$, where $1/\tau(\omega,T)$ is the scattering
rate and $\lambda(\omega,T)=m^*(\omega,T)/m_{b}-1$ is the
electron mass renormalization ($m_{b}$ is the bare band mass).
Note that the memory function definitions present in the
literature differ by a factor of $i$
\cite{GoetzePRB72,PuchkovJPCM96}; we follow the notation, where
$1/\tau(\omega,T)$ is given by the real part of $M(\omega,T)$.
In the case of a single band, $M(\omega,T)$ can be directly
extracted from the intraband optical conductivity via the
extended Drude formalism
\begin{eqnarray}\label{ExtDrude}
M(\omega,T)=\frac{1}{4\pi}\frac{\omega_{p}^2}{\sigma_{D}(\omega,T)}
+ i\omega.
\end{eqnarray}
For two bands, the situation is more complicated. The
band-dependent scattering has to be accounted for by taking
separate memory functions $M_{\sigma}(\omega,T)$ and
$M_{\pi}(\omega,T)$

\begin{eqnarray}\label{intraband}
\sigma_{D,\alpha}(\omega)=
\frac{1}{4\pi}\sum_{\beta=\sigma,\pi}
\frac{\omega_{p,\beta\alpha}^{2}}{M_{\beta}(\omega,T)-i\omega}
\mbox{, \ \  } (\alpha=a,c).
\end{eqnarray}

\noindent In fact, this complication is beneficial since the
scattering dissimilarity distinguishes optically the
contributions from different bands.

We assume that impurities/vacancies and phonons contribute
additively to the memory function (Matthiessen's rule):
\begin{equation}\label{memory1}
M_{\beta}(\omega,T) = \gamma_{\beta,\mbox{\tiny
imp}}+M_{\beta,\mbox{\tiny ph}}(\omega,T)\mbox{, \ \  }
(\beta=\sigma,\pi).
\end{equation}

\noindent The impurity scattering is strongly sample dependent
and frequency and temperature independent, in contrast to the
one due to lattice vibrations. The expression for the phonon
memory function at finite temperatures was obtained by Shulga
{\em et al.} \cite{ShulgaPC91}
\begin{equation}\label{memory2}
M_{\beta,\mbox{\tiny ph}}(\omega,T) = -2i\int_{0}^{\infty}\
\alpha^{2}_{\mbox{\tiny
tr}}F_{\beta}(\Omega)K\left(\frac{\omega}{2\pi
T},\frac{\Omega}{2\pi T}\right)d\Omega,
\end{equation}
\noindent which is a generalization of the famous $T=0$ formula
due to Allen \cite{AllenPRB71}. Here $\alpha^{2}_{\mbox{\tiny
tr}}F_{\beta}(\Omega)$ is the transport Eliashberg function,
and
\begin{eqnarray}
K(x,y) = \frac{i}{y} &+& \frac{y-x}{x}[\psi(1-ix+iy)-\psi(1+iy)]+\nonumber\\
& & \frac{y+x}{x}[\psi(1-ix-iy)-\psi(1-iy)]\nonumber
\end{eqnarray}

\noindent where $\psi(x)$ is the digamma function. The total
electron-phonon coupling strengths are given by the constants
\begin{eqnarray}
\lambda_{\mbox{\tiny
tr},\beta}=2\int_{0}^{\infty}\frac{\alpha^2_{\mbox{\tiny
tr}}F_{\beta}(\Omega)d\Omega}{\Omega}.
\end{eqnarray}

\noindent Eq. (\ref{memory2}) follows from the Kubo formula in
the weak-coupling limit, although it remains accurate even at
$\lambda_{\mbox{\tiny tr}}\sim 1$ \cite{FrankPrivate}. There
are several {\em ab initio} calculations of the Eliashberg
functions \cite{KongPRB01,ChoiPRB02,LiuPRL01} which all show a
strong coupling in the $\sigma$ bands and a moderate coupling
in the $\pi$ bands.

The memory function as determined from the experimental data by
Eq.(\ref{ExtDrude}) is proportional to the plasma frequency
squared. Therefore the strong underestimation of $\omega_{p}$
in the first optical experiments resulted in anomalously small
electron-phonon coupling constants \cite{MaksimovPRL02}. For
example, in Ref. \cite{TuPRL01} a value $\lambda\sim 0.13$ was
found, which is incompatible with the $T_{c}$ of 40 K without
invoking exotic coupling mechanisms \cite{MarsiglioPRL01}.

\begin{figure}[htb]

  \centerline{ \includegraphics[width=7cm]{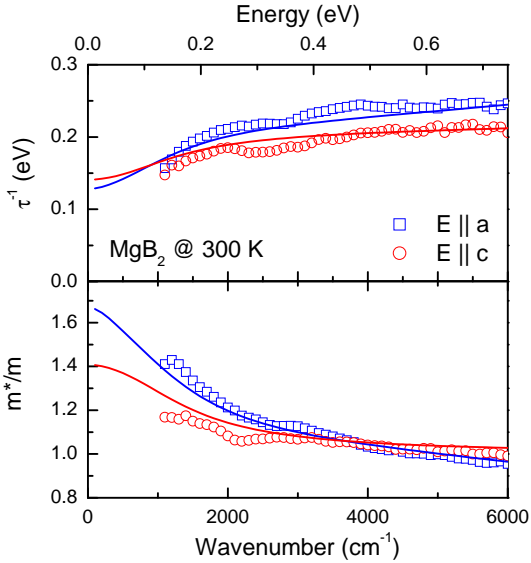}}
  \caption{Extended Drude analysis of the optical conductivity of
MgB$_2$ at 300 K along the in-plane and the c-axis directions.
The symbols are the data, the solid curves show a two-component
calculation. Adapted from Guritanu {\em et al.}
\cite{GuritanuPRB06}.}\label{FigExtdrudeAC}
\end{figure}

Fig.\ref{FigExtdrudeAC} shows $1/\tau(\omega)$ and
$m^*(\omega)/m_{b}$ at 300 K obtained by Guritanu et
al.\cite{GuritanuPRB06} by applying Eq. (\ref{ExtDrude}) to the
single crystal data using $\omega_{p,a}$ $\approx$
$\omega_{p,c}$ $\approx$ 6.3 eV. Since the plasma frequency of
cylindrical $\sigma$ bands is predicted by the LDA calculations
\cite{KortusPRL01,AntropovSHTS02} to be very small along the
$c$ axis (Fig.\ref{Wp}), the extended Drude parameters along
the c axis is almost entirely determined by the $\pi$ bands. In
contrast, the contributions from $\sigma$ and $\pi$ bands to
the in-plane properties are comparable, therefore both
scattering rate and mass renormalization for $E\parallel ab$
are effectively averaged over the two bands. This is why a
larger $m^*(\omega)/m_{b}$ and a faster growth of
$1/\tau(\omega)$ with photon energy for the in-plane
polarization indicate, without any modeling, that the
electron-phonon coupling in the $\sigma$ bands is stronger than
the one in the $\pi$ bands. One has to keep in mind though that
the $\sigma\rightarrow\sigma$ interband peak is predicted to be
at $\sim$ 0.3-0.5 eV for the in-plane polarization (see Section
\ref{Interband}) that may interfere with the extended Drude
analysis.

Kakeshita {\em et al.} \cite{KakeshitaPRL06} found that the
c-axis optical conductivity is much better described by a
single Drude peak with a frequency independent scattering rate
than the conductivity parallel to the ab plane. This agrees
well with the much weaker frequency dependence of
$1/\tau(\omega)$ for $E\parallel c$ than for $E\parallel ab$
\cite{GuritanuPRB06} (Fig.\ref{FigExtdrudeAC}).

A more quantitative comparison between experiment and theory
can be made using Eqs.(\ref{intraband}), (\ref{memory1}) and
(\ref{memory2}). Although not all parameters that enter these
formulas (four plasma frequencies, two impurity scattering
rates, and two Eliashberg functions) can be independently
derived from the data fit, the measured anisotropy of optical
spectra imposes tight constraints on them. At frequencies above
the phonon range the data are sensitive to the total coupling
constants but not to the spectral structure of the Eliashberg
functions. It was found  \cite{GuritanuPRB06} that the {\em ab
initio} calculated coupling constants ($\sim$ 1.1 for the
$\sigma$-bands and $\sim$ 0.55 for the $\pi$-bands
\cite{KortusPRL01,KongPRB01}) agree well with the spectral
data, if the quasi-2D nature of the $\sigma$ bands is assumed.

It appears that both the in-plane and c-axis infrared
reflectivity spectra are quite sensitive to the $\pi$-band
impurity scattering , allowing an accurate determination of
$\gamma_{\pi,\mbox{\tiny imp}}$ ($\gamma_{\pi,\mbox{\tiny
imp}}$ $\approx$ 85 meV for the single crystal from
Ref.\cite{GuritanuPRB06}). In contrast,
$\gamma_{\sigma,\mbox{\tiny imp}}$ cannot be reliably derived
from infrared spectra alone, because the scattering in the
$\sigma$ band is dominated by the electron-phonon interaction.
The ambiguity can be fixed by considering the temperature
dependence of the in-plane resistivity $\rho_{a}(T)$, for which
the ratio $\rho_{a}(300 K)/\rho_{a}(40 K)$ (RRR) is rather
sensitive to $\gamma_{\sigma,\mbox{\tiny imp}}$. In order to
match the experimental curve measured on a sample grown in the
same conditions that showes RRR $\approx$ 5,
$\gamma_{\sigma,\mbox{\tiny imp}}$ is about 7 times smaller
than $\gamma_{\pi,\mbox{\tiny imp}}$ had to be assumed. Such a
larger difference of the band scattering rates agrees with
Raman data \cite{QuiltyPRL03}. One should note that this
supports the hypothesis of Mazin {\em et al.} \cite{MazinPRL02}
that the $\sigma$ holes with wavefunctions strongly confined
within the boron planes should be scattered by typical lattice
defects (except the carbon atoms) much less than the $\pi$-band
carriers. This is due a larger probability of the defect
formation in the interstitials than inside the covalently
bonded boron sheets.

Kakeshita {\em et al.} \cite{KakeshitaPRL06} observed a strong
increase of the scattering rate in the $\sigma$ bands and a
significant but lesser increase in the $\pi$ bands with carbon
doping. This can be explained by the fact that C substitutes
boron atoms, affecting strongly the $\sigma$ band scattering. A
similar result was obtained by Di Castro {\em et al.}
\cite{DiCastroPRB06}, who found that carbon doping has a more
pronounced effect on the $\sigma$ band scattering than the Al
substitution.

\section{Interband transitions} \label{Interband}

The interband optical conductivity is determined by the band
dispersion and transition matrix elements
\cite{AntropovSHTS02,DresselGruner}

\begin{eqnarray}\label{SigIB}
\sigma_{1IB,\alpha}(\omega)&=&\frac{2\pi e^2}{V\omega
m^2}\sum_{k}\sum_{f,i\neq f}|\langle
kf|\nabla_{\alpha}|ki\rangle|^2
f(\epsilon_{ki})[1-f(\epsilon_{kf})]\nonumber\\
&\times&\delta(\epsilon_{kf}-\epsilon_{ki}-\hbar\omega)
\end{eqnarray}

\noindent where the indices $i(f)$ count bands of initial
(final) transition states, the factor 2 takes the different
spin orientations into account. The observed interband peaks
thus allow for an extra check of the band structure
calculations. The scattering processes smear the interband
spectra, which is often roughly accounted for by a spectral
convolution with a Gaussian.

\begin{figure}[htb]
  \includegraphics[width=9cm,clip=true]{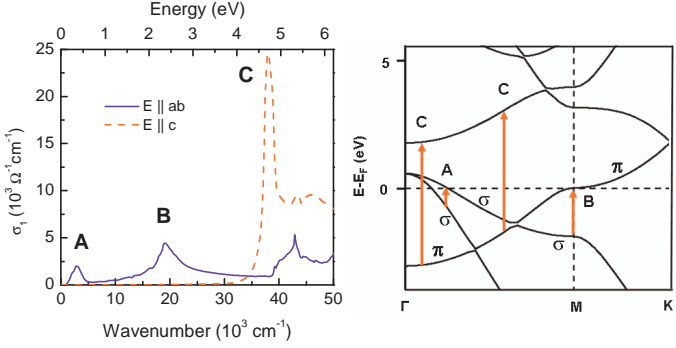}
  \caption{Left panel: the anisotropic interband optical conductivity of
  MgB$_{2}$ (adapted from Ref.\cite{KuzmenkoSSC02})
  obtained by {\em ab initio} calculations. Right panel: the band structure
  close to the Fermi level (adapted from Ref.\cite{MazinPC03}). The origin of
  the most prominent interband peaks A, B and C is show schematically by arrows.
  }\label{FigInterband}
\end{figure}

Several groups
\cite{RavindranPRB01,KuzmenkoSSC02,AntropovSHTS02,KuPRL02}
calculated $\sigma_{IB}(\omega)$ for MgB$_{2}$ all in
reasonable agreement with each other. Fig.\ref{FigInterband}
shows the calculated in-plane and c-axis interband
conductivities from Ref.\cite{KuzmenkoSSC02}. A remarkable
anisotropy is present: there is no sizeable optical intensity
along the c axis below 4 eV, whereas there are two peaks at
0.3-0.5 eV (marked as A) and 2.4 eV (marked as B) for
polarization parallel to the boron planes. Peak A corresponds
to the transition between two close $\sigma$ bands. Peak B is
due to a transition from the $\sigma$ band to the $\pi$ band
close to the M point, where a van Hove singularity strongly
enhances the density of states. The quasi-2D character of the
$\sigma$ bands explains why the transitions A and B are not
optically observed along the c axis. The c-axis conductivity
features a remarkably intense interband peak at about 5 eV,
which comes from a transition between almost parallel bands.

The $\sigma\rightarrow\pi$ transition (peak B) is observed in
Refs.\cite{GuritanuPRB06} and \cite{KakeshitaPRL06}. In the
ellipsometric study by Guritanu {\em et al.}
\cite{GuritanuPRB06}, this peak was found to be centered at 2.6
eV (Fig.\ref{FigDataAC}), which is slightly higher than the
theoretical value 2.4 eV \cite{KuzmenkoSSC02}. The mismatch in
the peak position suggests that the separation between
$\sigma$-band and $\pi$-bands is bigger than predicted by the
theory by about 200 meV. Interestingly a shift of the same
amount brings the results of the dHvA experiments
\cite{YellandPRL02} in MgB$_2$ close to theoretical predictions
\cite{RosnerPRB02}. Kakeshita {\em et al.}
\cite{KakeshitaPRL06} reported the interband transition even at
a higher energy of 2.8-2.9 eV (Fig.\ref{KakeshitaPRL06_Fig2})
which could be either due to the sample difference, or due to
the fact that the peak was close to the experimental frequency
cutoff of that study. No shift of the energy of this transition
with carbon substitution was observed.

\begin{figure}[htb]
  \centerline{\includegraphics[width=9cm]{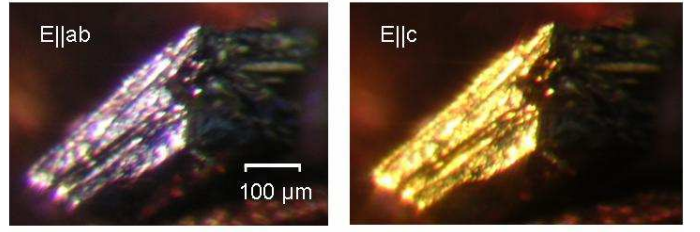}}

  \caption{Two photographs of the same crystal of MgB$_{2}$ (ac-plane) made in
polarized light: $E\parallel ab$ (left) and $E\parallel c$
(right). Adapted from Guritanu {\em et al.}
\cite{GuritanuPRB06}.}\label{FigColor}
\end{figure}

As a result of its large optical strength and the proximity to
the screened plasma frequency, the $\sigma\rightarrow\pi$
transition broadens the in-plane reflectivity plasma edge and
shifts it down by about 0.5 eV with respect to the one along
the c-axis (as was noticed in section \ref{Plasma}, the bare
plasma frequency appears to be almost isotropic!). A beautiful
manifestation of the plasma edge anisotropy is the multicolor
appearance of magnesium diboride \cite{GuritanuPRB06,LeePC03}.
The ac-oriented single crystals looks golden in light polarized
along the c-axis and blueish-silver for the in-plane
polarization (Fig.\ref{FigColor}). In the raw reflectivity
data, this transition causes a second pseudo plasma edge
(Fig.\ref{FigDataAC}), that was already seen in the first
single crystal spectra by Perucchi {\em et al.}
\cite{PerucchiPRL02}.

Less clear is the situation with the low-lying
$\sigma\rightarrow\sigma$ transition. According to the
calculations, it should be seen as a minute dip of about 1 \%
on the in-plane reflectivity \cite{GuritanuPRB06}. Di Castro
{\em et al.} reported such a feature at about 0.5 eV and found
that it shifts to lower energies with carbon and aluminum
doping. However, this effect is not obvious in other optical
studies, for example in
Refs.\cite{GuritanuPRB06,KakeshitaPRL06,TuPRL01}. More
experiments are needed to clarify this issue.

Unfortunately, the experimental range of the c-axis optical
studies \cite{GuritanuPRB06,KakeshitaPRL06} did not allow
seeing the 5 eV peak (peak C in Fig.\ref{FigInterband}). An
ultraviolet c-axis experiment is highly desirable to compare
its position with theoretical predictions.

\section{Far-infrared spectra: one gap or two?} \label{FIR}

Far-infrared transmission data of thin lead films
\cite{GloverPR56} showed the existence of the superconducting
gap before the advent of the BCS theory. Not surprisingly, many
groups examined the unusual superconductivity in MgB$_{2}$ with
far-infrared spectroscopy. Although infrared spectra do not
tell as directly as tunneling spectra about the electronic
density of states, the superconductivity induced correlations
have a dramatic effect on radiation absorption on the energy
scale of the gap. In an isotropic s-wave BCS superconductor,
the real part of the optical conductivity $\sigma_{1}(\omega)$
below $2\Delta$ vanishes at low temperatures, because photons
with a lower energy cannot break Cooper pairs. The inductive
component $\sigma_{2}(\omega)$ shows $\sim 1/\omega$ behavior
due to the formation of the condensate.

An early far-infrared grazing-incidence reflectivity
measurement by Gorshunov {\em et al.} \cite{GorshunovEPJB01} on
a porous polycrystalline sample of MgB$_{2}$ ($T_{c}$=39 K)
revealed a superconductivity-induced rise of reflectivity with
a maximum effect at about 3-4 meV, which was tentatively
attributed to the lowest value of $2\Delta$. Further
experiments on thin films
\cite{PimenovPRB02,JungPRB02,LeePRB02,KaindlPRL02,ZeleznyPC03,Lobo05},
thick films \cite{TuPRL01}, single crystals
\cite{PerucchiPRL02,DiCastroPRB06} and dense polycrystals
\cite{KuzmenkoSSC02,OrtolaniPRB05} provided much more detailed
and quantitative spectral information that in general supported
such assignment (Table \ref{TabGaps}). Note that this gap value
gives an anomalously small ratio $2\Delta/k_{B}T_{c}\sim 1-2$
as compared to the weak-coupling BCS result of 3.52. However,
it agrees well with the smallest gap observed in the $\pi$ band
by tunneling spectroscopy \cite{IavaronePRL02} and ARPES
\cite{SoumaNature03}.

\begin{table}
\caption{A summary of the superconducting gap values of
MgB$_{2}$ extracted from different far-infrared experiments.
Values marked by * are determined from the single-gap fit, by
** - from the maximum of $R_{s}/R_{n}$, by *** - from the onset
of absorption. }
\begin{center}
\begin{tabular}{ccccccc}
\hline
sample & refl/tran & $T_{c}$ (K) & $2\Delta$ (meV) & $\frac{2\Delta}{k_{B}T_{c}}$ & Ref.\\
\hline
thin film      & tran & 33   & 5.2*  & 1.8     & \cite{JungPRB02}\\
thin film      & tran & 30.5 & 5***  & 1.9     & \cite{KaindlPRL02}\\
thin film      & refl & 32   & 3***  & 1.1     & \cite{PimenovPRB02}\\
thin film      & refl & 30   & 5.2*  & 2.0     & \cite{Lobo05}\\
thin film      & refl & 35   & 5***  & 1.6     & \cite{ZeleznyPC03}\\
polycrystal    & refl & 39   & 3-4** & 0.9-1.2 & \cite{GorshunovEPJB01}\\
polycrystal    & refl & 39   & 4.8*  & 1.4     & \cite{OrtolaniPRB05}\\
thick film     & refl & 39.6 & 5***  & 1.5     & \cite{TuPRL01}\\
mosaic         & refl & 38   & 4***  & 1.2     & \cite{PerucchiPRL02}\\
single crystal & refl & 38.5 & 4*    & 1.2     & \cite{DiCastroPRB06}\\
\hline
\end{tabular}
\end{center} \label{TabGaps}
\end{table}

While a gap-like feature at an energy scale of the order of the
smaller gap $\Delta_{\pi}$ was universally seen, no clear
evidence of the larger gap $\Delta_{\sigma}$ ($\sim$ 7 meV
according to tunneling \cite{IavaronePRL02}, ARPES
\cite{SoumaNature03} and Raman \cite{QuiltyPRL02} data) was
found in the far-infared experiments. This is perhaps the most
puzzling and intriguing issue raised by the infrared
spectroscopy of magnesium diboride.

\begin{figure}[htb]
\centerline{
\includegraphics[width=9cm]{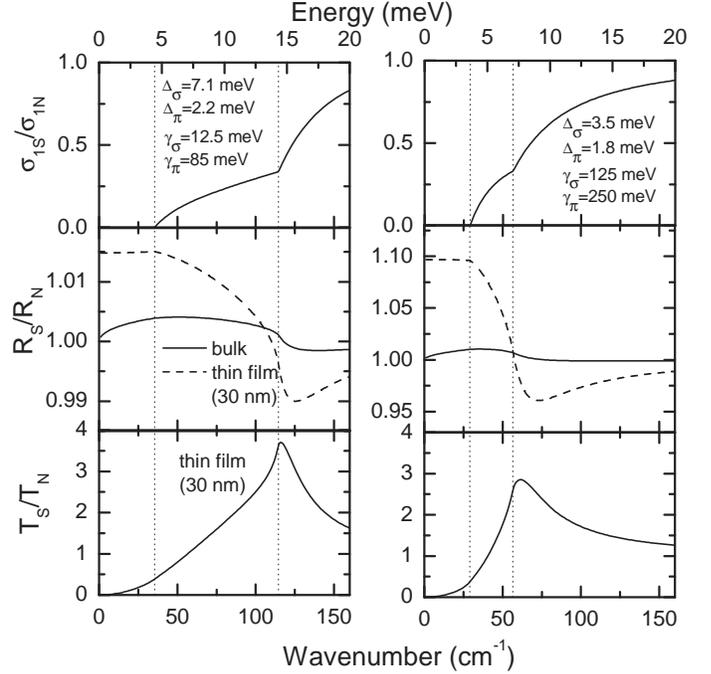}}
  \caption{A simulation of the far-infrared properties of a two-gap BCS
  superconductor using the Mattis-Bardeen formalism extended to arbitrary purity \cite{ZimmermannPC91}.
  The two cases are considered (left vs. right panels).
  The values of $\gamma$ and $\Delta$ for $\sigma$ and $\pi$ bands are given on the top panels. In both cases,
  $\omega_{p,\sigma a}$=4.1 eV, $\omega_{p,\pi a}$=6.9 eV. The vertical lines denote
  $2\Delta_{\pi}$ and $2\Delta_{\sigma}$.}\label{Firmodel}
\end{figure}

To simplify the comparison of different measurements, we begin
with a basic simulation of the expected effect of the two-gap
structure on the in-plane infrared spectra of MgB$_{2}$,
similar to the one made by Lee {\em et al.} \cite{LeePRB02} and
Lobo {\em et al.} \cite{Lobo05}. We assume that $\sigma$ and
$\pi$ charge carriers respond independently to the external
radiation; the condition that justifies using
Eq.(\ref{SigSum}). Each component is characterized by its own
in-plane plasma frequency $\omega_{p,a}$, scattering rate
$\gamma$ (which is mostly due to impurities at $T$ $<$ 40 K)
and superconducting gap $\Delta$. We use the Mattis-Bardeen
formalism \cite{MattisPR58} for an s-wave isotropic BCS
superconductor extended to the case of finite impurity
scattering by Zimmermann {\em et al.} \cite{ZimmermannPC91}.
The standard Fresnel equations are employed.

Figure \ref{Firmodel} shows the superconducting to normal-state
ratios $f_{S}/f_{N}=f(0)/f(T_{c})$ for the real part of the
optical conductivity (top), the reflectance of a bulk sample
(middle, solid line), the one of a thin film with the thickness
of 30 nm (middle, dashed line) and for the transmission of the
same thin film (bottom). In the first simulation (left panels),
it is assumed that the gap values $\Delta_{\sigma}$ = 7.1 meV
and $\Delta_{\pi}$ = 2.2 meV are the same as in the tunneling
spectra \cite{IavaronePRL02}. The scattering rates
$\gamma_{\sigma}$ = 12.5 meV and $\gamma_{\pi}$ = 85 meV are
taken from the optical study \cite{GuritanuPRB06}. The plasma
frequencies are borrowed from the LDA calculation
\cite{KortusPRL01} which are close to the recent optical
results \cite{GuritanuPRB06,KakeshitaPRL06} as discussed in
Section \ref{Plasma}. We ignored the effective mass
renormalization due to electron-phonon interaction which does
not qualitatively change the results of this simulation. The
conductivity $\sigma_{1}(\omega)$ shows two independent
thresholds at $2\Delta_{\pi}$ and $2\Delta_{\sigma}$. The
reflectivity ratio is slightly less than 1 for $\omega >
2\Delta_{\sigma}$ showing a minimum somewhat above
$2\Delta_{\sigma}$, but at lower energies it suddenly grows and
saturates (for a thin film) or shows a broad maximum (for a
bulk sample) at about $2\Delta_{\pi}$. The film reflectance
structure becomes quite sharp if the thickness is less than the
penetration depth. This generally makes thin films experiments
more sensitive to the gap value than the ones made on bulk
samples. The film transmittance shows a pronounced peak
centered near $2\Delta_{\sigma}$, while the $\pi$ gap only
slightly modifies the curve shape without producing any sharp
spectral features.

\begin{figure}[htb]
\centerline{
\includegraphics[width=8cm]{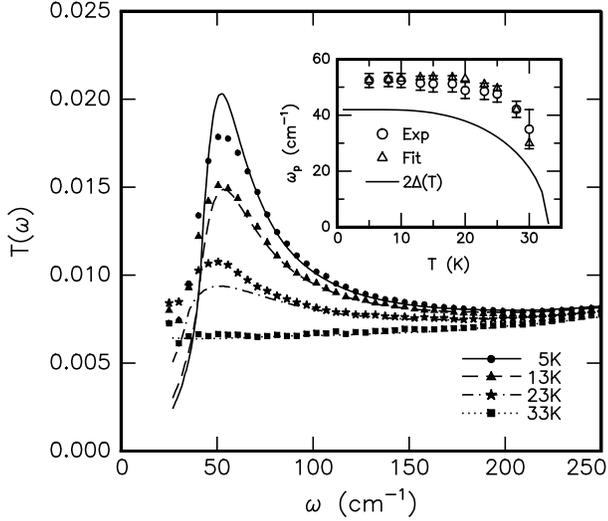}}
  \caption{Far-infrared transmittance of a thin c-axis oriented MgB$_{2}$
  film ($T_{c}$ $\approx$ 33 K). Solid line is fit by a one-gap model. The inset shows the
  temperature dependence of the peak position and $2\Delta$.
  Adapted from Jung {\em et al.} \cite{JungPRB02}. }\label{FigJungPRB02}
\end{figure}

Jung {\em et al.} \cite{JungPRB02} measured the far-infrared
transmittance of a thin ($\sim$ 50 nm) MgB$_{2}$ film ($T_{c}$
= 33 K) with the c axis perpendicular to the surface
(Fig.\ref{FigJungPRB02}). The transmittance in the
superconducting state showed a peak at around 6.5 meV at 5 K.
The best fit assuming a single gap value was achieved with
$2\Delta$ $\approx $ 5.2 meV. The temperature dependence of the
gap was close to the prediction of the BCS theory. However, the
peak position is about two times (!) lower than the prediction
of the two-gap model based on the tunneling result
(Fig.\ref{Firmodel}). To be more accurate, one needs to correct
for a reduced $T_{c}$ by taking a smaller value of
$\Delta_{\sigma}$. From the solution of full Eliashberg
equations, Dolgov {\em et al.} \cite{DolgovPRB05} found that
$\Delta_{\sigma}$ should be about 5.5 meV for $T_{c}$=33 K,
which is close to the recent observations on neutron-irradiated
MgB$_{2}$ samples \cite{PuttiPRL06}. However this correction
seems to be by far insufficient to fix the mismatch.

\begin{figure}[htb]
\centerline{
\includegraphics[width=6cm]{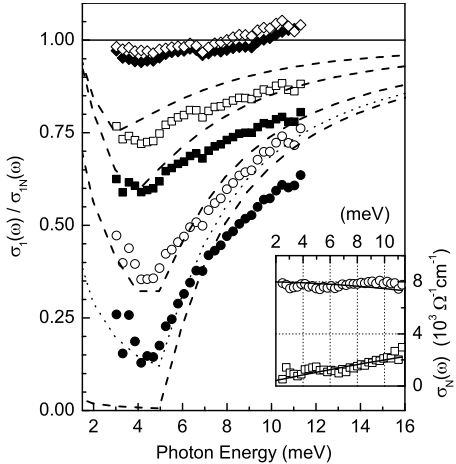}}
  \caption{Real part of conductivity $\sigma_{1}(\omega)$ of thin MgB$_{2}$ film ($T_{c}$ = 30.5 K) normalized to its normal state value at 40 K.
  From top to bottom: 50, 30, 27, 24, 17.5 and 6 K. Solid curves obtained by Mattis-Bardeen calculations (27, 24, 17.5 and 6
  K). Inset: real (circles) and imaginary (squares)
part of normal state conductivity at 40 K, along with a Drude
calculation (lines).
  Adapted from Kaindl {\em et al.} \cite{KaindlPRL02}.}\label{FigKaindlPRL02}
\end{figure}

Kaindl {\em et al.} \cite{KaindlPRL02} reported
$\sigma_{1}(\omega)$ and $\sigma_{2}(\omega)$ of two MgB$_{2}$
films ($T_{c}$ = 30 and 34 K) directly obtained by terahertz
time domain spectroscopy in transmission geometry. The
transmission peak was at about 7 meV for both films, which is
close to the observation in Ref.\cite{JungPRB02}. A strong
depletion of absorption was observed below $\sim$ 5 meV at 6 K,
although the data do not exclude that the full conductivity
suppression is at somewhat lower energy
(Fig.\ref{FigKaindlPRL02}).

\begin{figure}[htb]
\centerline{
\includegraphics[width=6cm]{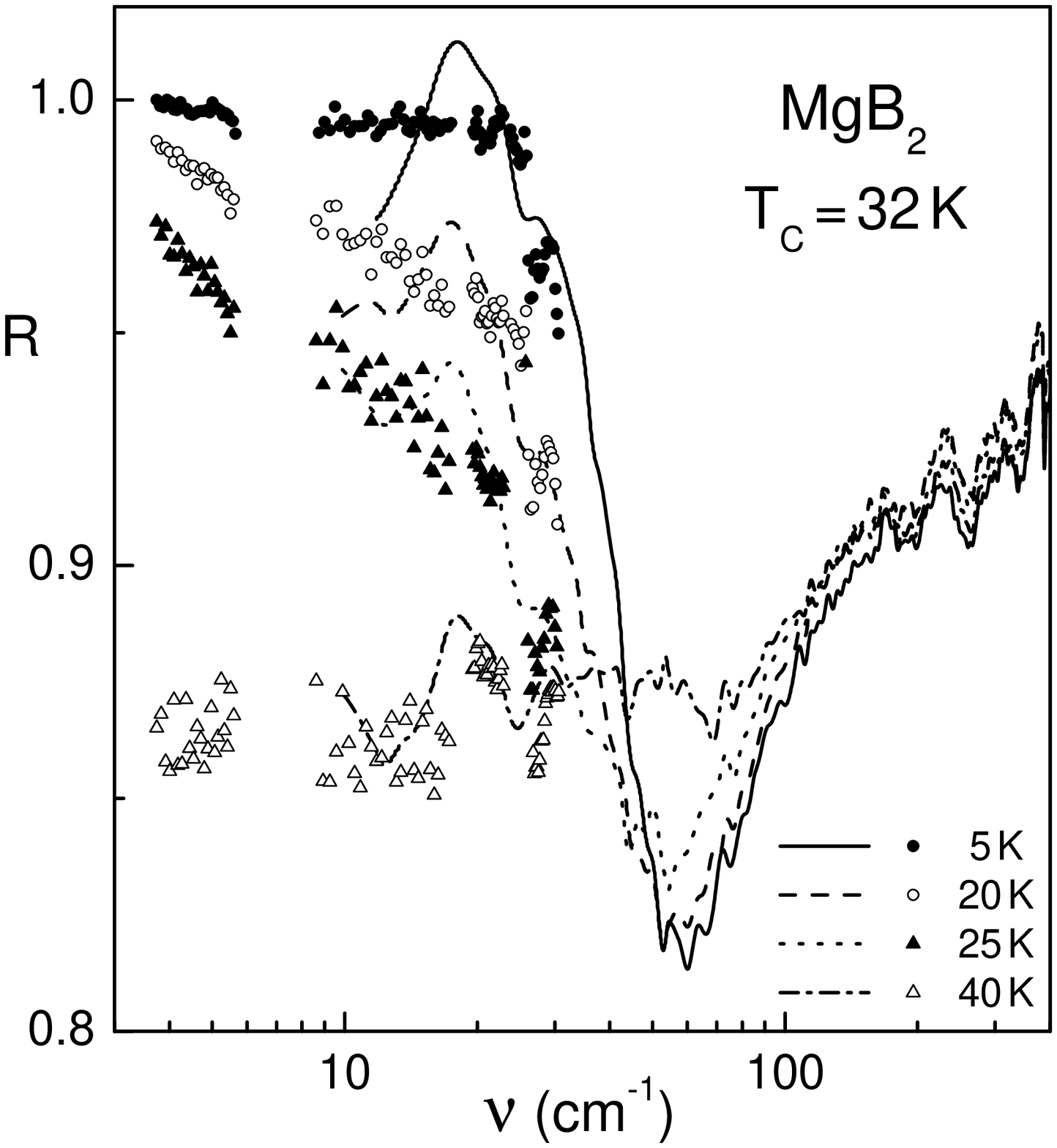}}
  \caption{Far-infrared (lines) and submillimeter (dots) reflectance of a thin MgB$_{2}$ film.
  Adapted from Pimenov {\em et al.} \cite{PimenovPRB02}.
  }\label{FigPimenovPRB02}
\end{figure}

Pimenov {\em et al.} \cite{PimenovPRB02} obtained the
reflectance of a thin MgB$_{2}$ film with a similar $T_{c}$ (32
K) (Fig.\ref{FigPimenovPRB02}). The spectral range was extended
down to 0.5 meV by using a quasioptical backward wave
oscillator technique which gives directly $\sigma_{1}$ and
$\sigma_{2}$. The onset of absorption was observed at about 3
meV, which was ascribed to the value of $2\Delta$. Although
this number is smaller than 4.2 meV found in
Ref.\cite{JungPRB02}, one should realize that the way of
determination is different. In fact, a pronounced reflectivity
dip is observed in Ref.\cite{PimenovPRB02} at $\sim$ 7 meV,
which is close to the transmittance peak from Refs.
\cite{JungPRB02} and \cite{KaindlPRL02}. A similar result was
obtained by Lobo {\em et al.}\cite{Lobo05} on a c-axis oriented
film with $T_{c}$ $\approx$ 30 K. This peak-dip match suggests
the mutual consistency of different thin film studies.

While the straightforward application of the model assuming a
large gap separation clearly fails to describe the existing
infrared spectra, one may wonder whether these data are still
compatible with the presence of two, less distinct, effective
gaps. Lobo {\em et al.} \cite{Lobo05} found that a two-gap
model with $\Delta_{\pi}$ = 1.8 meV and $\Delta_{\sigma}$ = 3.5
meV improves the fit of the reflectivity curve compared to the
best fit by the one-gap model with $2\Delta$ = 2.6 meV,
although the improvement was not sufficient to conclude that
two gaps must exist. In Fig.\ref{Firmodel} (right panel) a
second simulation with the same formulas is shown where the
parameters are changed in two ways: (i) the gaps are reduced to
$\Delta_{\pi}$ = 1.8 meV and $\Delta_{\sigma}$ = 3.5 meV as in
Ref.\cite{Lobo05}, which nevertheless leaves them quite
distinct and (ii) the sample is assumed to be far in the dirty
limit ($\gamma_{\sigma}$ = 125 meV and $\gamma_{\pi}$ = 250
meV), which is perhaps a better description for thin films as
compared to single crystals. This model qualitatively matches
both reflection \cite{PimenovPRB02,Lobo05} and transmission
\cite{JungPRB02,KaindlPRL02} spectra.

Single crystals of magnesium diboride have an advantage of
higher $T_{c}$ and smaller residual resistivity as compared to
thin films. The first single-crystal optical measurement of
MgB$_{2}$ was made by Perucchi {\em et al.}
\cite{PerucchiPRL02} on a mosaic of ab-oriented samples
($T_{c}$=38 K). The spectra were taken down to 3 meV while
changing temperature down to 1.6 K and applying magnetic fields
up to 7 T, which allowed a fully optical determination of the
temperature dependence of the c-axis critical field $H_{c2}$
($\sim$ 5 T at lowest 1.6 K). The authors \cite{PerucchiPRL02}
observed the absorption threshold at about $\sim$ 4 meV, which
is close to the observation on thin films
\cite{KaindlPRL02,PimenovPRB02}. Di Castro {\em et al.}
\cite{DiCastroPRB06} measured the ratio $R_{S}/R_{N}$ on
separate ab-oriented single crystals of MgB$_{2}$ ($T_{c}$=38.5
K) using the synchrotron radiation between 3 and 12 meV to
compensate the loss of signal due to the small sample size.
They observed a relative reflectivity increase below $T_{c}$,
which is much smaller in size as compared to the one from
Ref.\cite{PerucchiPRL02}, but much more close to the
expectation according to the Mattis-Bardeen-Zimmermann
calculation (Fig.\ref{Firmodel}). The rise of reflectivity was
observed only below 5 meV at zero doping and 8 meV for a carbon
substituted sample, which is much lower than the prediction of
the model that assumes a large value of $\Delta_{\sigma}$
(Fig.\ref{Firmodel}, left panels). The reason for a
significantly different size of the superconductivity induced
increase of reflectance in Refs.\cite{PerucchiPRL02} and
\cite{DiCastroPRB06} is not fully clear, although it may be
related to mosaic effects in the first measurement or
diffraction effects due to a small sample size in the second
one. Further experiments on single crystals may clarify this
issue.

Let us now speculate on the reasons for the failure of the
straightforward two gap description of the infrared spectra in
MgB$_{2}$. One possibility is that the assumption about two
independent responses of $\sigma$ and $\pi$ bands has to be
abandoned. Another weakness of the modeling of
Fig.\ref{Firmodel} is the assumption about the local
electromagnetic response and the validity of the Fresnel
equations. This would be well justified in the dirty limit,
which is likely to be applicable to the $\pi$ bands, but not
necessarily to the $\sigma$ bands. The last possibility is that
there {\em are} two gaps in the infrared spectra, but they are
effectively reduced for some reason with respect to the ones
observed by other probes. Certainly more theoretical and
experimental research is needed to solve this paradox.

To get the full picture of the electromagnetic response of
MgB$_{2}$ in the superconducting state one would have to
complement the discussed {\em spectroscopic} data with the {\em
temperature} dependence of $\sigma_{1}$ and $\sigma_{2}$
(usually expressed via the penetration depth $\lambda$) as
measured by microwave \cite{JinPRL03,LeePRB05,FletcherPRL05}
and submillimeter \cite{ProninPRL01} techniques. Interestingly,
the structure of the so-called coherence peak is highly
anomalous for a single gap superconductor, revoking the
explanation in terms of two gaps (some of these measurements
are reviewed in Ref. \cite{GolubovPRB02}). Optical studies of
photoexcited quasiparticle dynamics \cite{DemsarPRL03,Lobo05}
provide another useful piece of information. The discussion of
these latter experiments is unfortunately beyond the current
review.

\section{Conclusions}
\label{Conclusions}

Many optical properties of magnesium diboride now appear to be
consistent with the {\em ab initio} electron and phonon band
calculations. The long standing problem of an apparently small
plasma frequency is solved due to the recent measurements on
single crystals: it is close to the theoretical prediction of 7
eV and is almost isotropic (which should be regarded, in fact,
as a coincidence, since the plasma frequency of the $\sigma$
bands is very small along the c axis). The c-axis plasma
frequency strongly decreases with carbon doping, in contrast to
the one parallel to the boron planes. The observed energy of
the $\sigma\rightarrow\pi$ interband transition (2.6 eV)
suggests that the LDA calculations underestimate the separation
between $\sigma$ and $\pi$ bands by about 200 meV, in agreement
with the de Haas-van Alphen experiments. Optical data in a
combination with DC resistivity curves show that the $\sigma$
bands are characterized by a stronger electron-phonon coupling
but a smaller impurity scattering as compared to the $\pi$
bands.

Far-infrared spectra clearly show a gap-like onset of
absorption indicating $2\Delta$ = 3-5 meV and an anomalously
small ratio $2\Delta/k_{B}T_{c}$=1-2. This value likely refers
to the gap in the $\pi$ bands. A serious problem which remains
is the absence of a clear far-infrared signature of the large
gap seen in other experiments in MgB$_{2}$. Before this issue
is resolved, our understanding of two-gap superconductivity can
hardly be considered as a complete.

\section{Acknowledgments}

The author is grateful to V. Guritanu, D. van der Marel, O.V.
Dolgov, J. Kortus, I.I. Mazin, J. Karpinski, R.P.S.M. Lobo, M.
Ortolani, L. Degiorgi, S. Tajima and T. Kakeshita for
discussions and to E. van Heumen and V. Guritanu for a help in
the preparation of this manuscript. The financial support was
provided by the Swiss National Science Foundation through the
National Center of Competence in Research "Materials with Novel
Electronic Properties-MaNEP".

\end{document}